\documentclass[prb,twocolumn,amsmath,amssymb,showkeys]{revtex4}

\usepackage{graphicx}
\usepackage{bm}

\newcommand{\lang}{\left\langle}
\newcommand{\rang}{\right\rangle}
\newcommand{\ssum}{\textstyle\sum}

\begin{document}

\title{Stochastic actions for diffusive dynamics: Reweighting, sampling, and minimization}

\author{Artur B. Adib}
\email{adiba@mail.nih.gov}
\affiliation{
Laboratory of Chemical Physics, NIDDK, National Institutes of Health, Bethesda, Maryland 20892-0520, USA
}

\date{\today}

\begin{abstract}
In numerical studies of diffusive dynamics, two different action functionals are often used to specify the probability distribution of trajectories, one of which requiring the evaluation of the second derivative of the potential in addition to the force. Here it is argued that both actions are equivalent prescriptions for the purposes of reweighting and sampling trajectories, whereas the most probable path is more generally given by the global minimum of the action involving the second derivative term. The answer to this apparent paradox lies in the non-differentiable character of Brownian paths, as well as in the ``entropy'' associated with a given trajectory.
\end{abstract}


\maketitle


\section{Introduction}

The present paper is concerned with the probability distribution of diffusive trajectories, particularly in discrete form as suited for numerical computations. In its simplest version, the problem is that of specifying the probability that a Brownian particle moving according to the overdamped Langevin equation\cite{zwanzig01}
\begin{equation} \label{langevin}
  \dot{x}(t) = \frac{F(x(t))}{\zeta} + \sqrt{2 D} \, \mathcal{R}(t)
\end{equation}
will follow a particular path from an initial to a final position in a given time. Here $F(x) = - U'(x)$ is the force acting on the particle, $\zeta$ is the friction coefficient, $D$ is the diffusion constant, and $\mathcal{R}(t)$ is a random force term satisfying the white noise relation $\lang \mathcal{R}(t) \mathcal{R}(t') \rang = \delta(t-t')$. The temperature in energy units, $kT$, is related to the above parameters via the Einstein relation, $D = kT/\zeta$. 

Unlike their deterministic counterparts, stochastic differential equations such as Eq.~(\ref{langevin}) remain ambiguous until one specifies the discretization rule for evaluating the various time-dependent quantities.\cite{gardiner-book} Two typical rules are the Ito and Stratonovich discretizations, in which $F(x(t))$ is evaluated at the beginning and at the mid-point of the time-interval, respectively (see Eq.~(\ref{ito}), for example). In the continuum limit, it is well known that both choices lead to (a) the same Fokker-Planck equation (provided $D$ is independent of $x$; cf. Eq.~(\ref{FP})),\cite{gardiner-book} and (b) the same path integral formula for the propagator, expressed in terms of the so-called action functional (cf. Eqs.~(\ref{path-int})-(\ref{S})).\cite{hunt-ross81}

The above results are consistent with the intuitive expectation that the measurable statistical properties of a Brownian particle should be insensitive to the particular stochastic calculus chosen to describe its motion. Nonetheless, an ambiguity of practical character arises when deriving a discrete form for the probability distribution of trajectories, i.e. the discretized action: Depending on the type and the stage at which one carries out the discretization, the ensuing action can take on different functional forms (cf. Sec.~\ref{sec:background}). In particular, in previous numerical studies concerned with sampling,\cite{pratt86,dellago98a,doniach01} reweighting,\cite{jarzynski99-NucPhys,zuckerman99,zuckerman00,xing06} and optimizing\cite{dellago98b,elber00,orland06} diffusive trajectories, one typically finds two different expressions for the stochastic action, and no consensus seems to exist as to which form should be used in a given application.\cite{zuckerman00,elber00,orland06}

It is the purpose of this paper to shed some light on this issue. Of course, the scope of this problem goes well beyond the prototypical case of a Brownian particle moving in a field of force; for example, chemical reactions in solution are typically described by effective equations of motion that are straightforward multi-dimensional extensions of Eq.~(\ref{langevin}).\cite{zwanzig01,nitzan-book} A specific example of increasing interest in recent years is the microscopic description of the mechanisms underlying protein folding, which often requires the characterization of diffusive paths connecting unfolded and folded states.\cite{doniach01,orland06} In these and similar cases, it is crucial to understand and resolve the aforementioned ambiguity. The next section offers a brief introduction to this problem, followed by specific applications to reweighting, sampling, and minimization in Sec.~\ref{sec:applications}.

\section{Two actions} \label{sec:background}

The central quantity of this work is the relative probability of observing a given diffusive trajectory connecting an initial ($x_i$) to a final ($x_f$) configuration. In the continuum limit, this probability is directly available from the path integral expression for the propagator, $P(x_f|x_i;t)$.\cite{chaichian-book01} For the dynamics in Eq.~(\ref{langevin}), regardless of the discretization adopted, one obtains\cite{hunt-ross81}
\begin{equation} \label{path-int}
  P(x_f|x_i;t) = \int_{x(0)=x_i}^{x(t)=x_f} \mathcal{D}x(s) \, e^{-S[x(s)]/2D},
\end{equation}
where
\begin{equation} \label{S}
  S[ x(s) ] = \frac{\Delta U}{\zeta} + \frac{1}{2} \int_0^t \! ds \! \left[ \dot{x}^2 + \left( \frac{F}{\zeta} \right)^2 + \frac{2D}{\zeta} F' \right]
\end{equation}
is the so-called action functional. Here $\Delta U \equiv U(x_f) - U(x_i)$, and the term involving the formal time derivative of $x$, viz. $\int \! ds\, \dot{x}^2$, is a shorthand for the free diffusion contribution to the path-integration measure, cf. Eq.~(\ref{S2}) (recall that Brownian trajectories are non-differentiable).\cite{chaichian-book01} Equation (\ref{path-int}) shows that each trajectory contributes to the propagator with a weight ultimately dictated by its action. This will be the object of interest in the remainder of the paper.

\subsection{Derivation}

Although in the continuum limit the action is given unambiguously by the above expression, different discrete forms can be obtained depending on how and when one discretizes the problem. Here I will follow the (implicit or explicit) argument that often appears in the literature to justify the use of two particular discrete actions. The first arises by performing the discretization at the level of Eq.~(\ref{langevin}) using the Ito rule (which incidentally leads to the algorithm traditionally adopted for the numerical generation of Brownian trajectories\cite{allen-tildesley87}), namely
\begin{equation} \label{ito}
  x_{n+1} = x_n + \frac{\Delta t}{\zeta} F_n + \sqrt{2 D \, \Delta t} \, R_n,
\end{equation}
where $\Delta t$ is the time step, $F_n \equiv F(x_n)$, and $R_n$ is a Gaussian random variable of unit variance. Because the distribution $P(R_n) = e^{-R_n^2/2}/ \sqrt{2\pi} $ is known and the force only depends on the previous position $x_n$, the short-time conditional probability $P(x_{n+1}|x_n; \Delta t)$ can be straightforwardly obtained by changing variables from $R_n$ to $x_{n+1}$, which in this particular case comes with unit Jacobian. The corresponding action, which comes from the definition
\begin{equation}
  P(x_N|x_{N-1};\Delta t) \cdots P(x_1|x_0;\Delta t) \equiv e^{-S_1[x_0\cdots x_N]/2D}
\end{equation}
up to a path-independent factor, is thus given by
\begin{equation} \label{S1}
  S_1[x_0 \cdots x_N] = \frac{\Delta t}{2} \sum_{n=0}^{N-1} \left( \frac{x_{n+1}-x_n}{\Delta t} - \frac{F_n}{\zeta} \right)^2,
\end{equation}
where $x_0 = x_i$ and $x_N = x_f$. Note that this dictates the {\em exact} probability of a path generated by Eq.~(\ref{ito}) for any finite $\Delta t$. On the other hand, a direct discretization at the level of Eq.~(\ref{S}) gives the action
\begin{multline} \label{S2}
  S_2[x_0 \cdots x_N] = \frac{\Delta U}{\zeta} + \frac{\Delta t}{2} \sum_{n=0}^{N-1} \left[ \left( \frac{x_{n+1}-x_n}{\Delta t} \right)^2 + \right. \\
  + \left. \left( \frac{F_n}{\zeta} \right)^2 + \frac{2D}{\zeta} F_n' \right].
\end{multline}
Aside from the formal $\dot{x}^2$ term correctly expressed above, Eq.~(\ref{S}) involves ordinary integrals only, and hence is insensitive to the particular discretization rule adopted in Eq.~(\ref{S2}).

\subsection{When are the actions equivalent?} \label{sec:equivalence}

The two actions derived above are manifestly different; in particular, the presence of the derivative of the force makes $S_2$ especially inconvenient for numerical implementations, which presumably explains why $S_1$ is more widely adopted in the literature\cite{pratt86,dellago98a,doniach01,jarzynski99-NucPhys} (see also Ref.~\onlinecite{predescu07} for a new action functional without this term). Nonetheless, the very validity of $S_1$ as the action for diffusive problems has been questioned by some authors.\cite{zuckerman00,orland06} The goal of this section is to clarify the (non-)equivalence between these two actions.

To begin the analysis, it is instructive to first consider an obvious situation where the actions are {\em not} equivalent, namely, when they are evaluated in the continuum limit over {\em differentiable} trajectories. In this case, Eq.~(\ref{S1}) becomes
\begin{equation} \label{S1cont}
  S_1[x(s)] = \frac{\Delta U}{\zeta} + \frac{1}{2} \int_0^t ds \left[ \dot{x}^2 + \left( \frac{F}{\zeta} \right)^2 \right],
\end{equation}
where the cross-term $\dot{x} F/\zeta$ was integrated by parts to yield the surface term $\Delta U/\zeta$. Comparing this result with the analogous limit of $S_2$, which is repeated here for the convenience of the reader (cf. Eq.~(\ref{S})), 
\begin{equation} \label{S2cont}
  S_2[ x(s) ] = \frac{\Delta U}{\zeta} + \frac{1}{2} \int_0^t \! ds \! \left[ \dot{x}^2 + \left( \frac{F}{\zeta} \right)^2 + \frac{2D}{\zeta} F' \right]
\end{equation}
we see that these two expressions differ by a non-trivial term whose magnitude is dictated by the ratio $D/\zeta$; i.e., the actions are generally not equivalent when evaluated over differentiable functions (the case where $F'$ is constant is an exception, e.g. the harmonic potential case considered by Onsager and Machlup\cite{onsager53a}).

\begin{figure}
\begin{center}
\includegraphics[width=240pt]{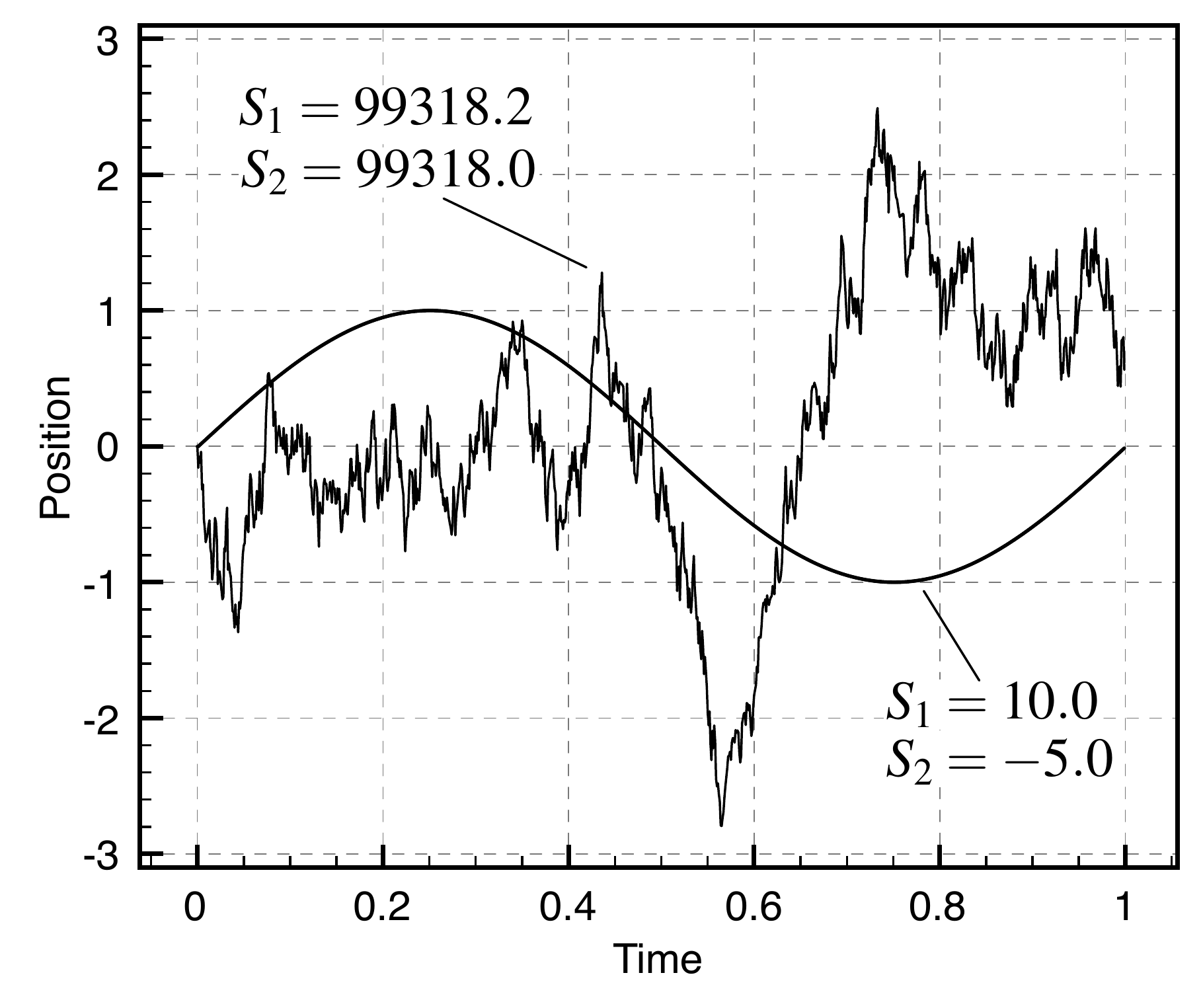} 
\caption{Illustrative values of the two discrete actions $S_1$ and $S_2$, Eqs.~(\ref{S1}) and (\ref{S2}) respectively, evaluated over a typical trajectory of a Brownian particle moving in a quartic potential (noisy curve, see below), as well as over an arbitrary differentiable function, $\sin(2 \pi t)$ (smooth curve). Note that the numeric values of the actions agree to within five significant digits when evaluated over the Brownian trajectory, while they differ by as much as their own magnitude in the case of the smooth curve. The actions were evaluated using the quartic potential $U(x) = x^4/4$, and the parameters $\Delta t = 10^{-4}$, $D = 10$, and $\zeta = 1$. The Brownian path was generated by iterating the algorithm in Eq.~(\ref{ito}) with the same potential and parameters. 
}
\label{fig:S1S2traj}
\end{center}
\end{figure}

Consider now the case where the actions are evaluated over Brownian paths, which are non-differentiable and probabilistic in nature. The continuum limit of Eq.~(\ref{S1}) then leads to a stochastic integral, which in general does not satisfy the usual relations from differentiable calculus. Indeed, in the limit of infinitely many intervals, the cross term that gave rise to the surface term above becomes an Ito integral, i.e.
\begin{equation} \label{ito1}
  \lim_{\substack{N\to \infty \\ \Delta t \to 0 }} \sum_{n=0}^{N-1} (x_{n+1} - x_n) F_n \equiv \text{(I)} \int_{x_i}^{x_f} \! dx \, F(x),
\end{equation}
which satisfies the (probabilistic) equality\cite{gardiner-book} (see also Appendix)
\begin{equation} \label{ito2}
  \text{(I)} \int_{x_i}^{x_f} \! dx \, f'(x) = \Delta f - D \int_{0}^t \! ds \, f''(x(s)).
\end{equation}
Here $D$ is the diffusion constant associated with the trajectory $x(t)$, $f(x)$ is an arbitrary differentiable function, and $\Delta f = f(x_f) - f(x_i)$. Note that in the zero-noise limit ($D \to 0$) one recovers the fundamental theorem of differentiable calculus; for any finite $D$, however, this result allows one to replace Ito integrals by ordinary time-integrals. Consequently, when expressed in terms of ordinary time-integrals only, the limiting form of $S_1$ acquires a term proportional to the second-derivative of the potential in addition to the terms shown in Eq.~(\ref{S1cont}), correctly recovering Eq.~(\ref{S2cont}). This shows the (probabilistic) equivalence between $S_1$ and $S_2$ in the continuum limit when evaluated over Brownian trajectories with same diffusion constant as that appearing in the definition of $S_2$.

\begin{figure}
\begin{center}
\includegraphics[width=240pt]{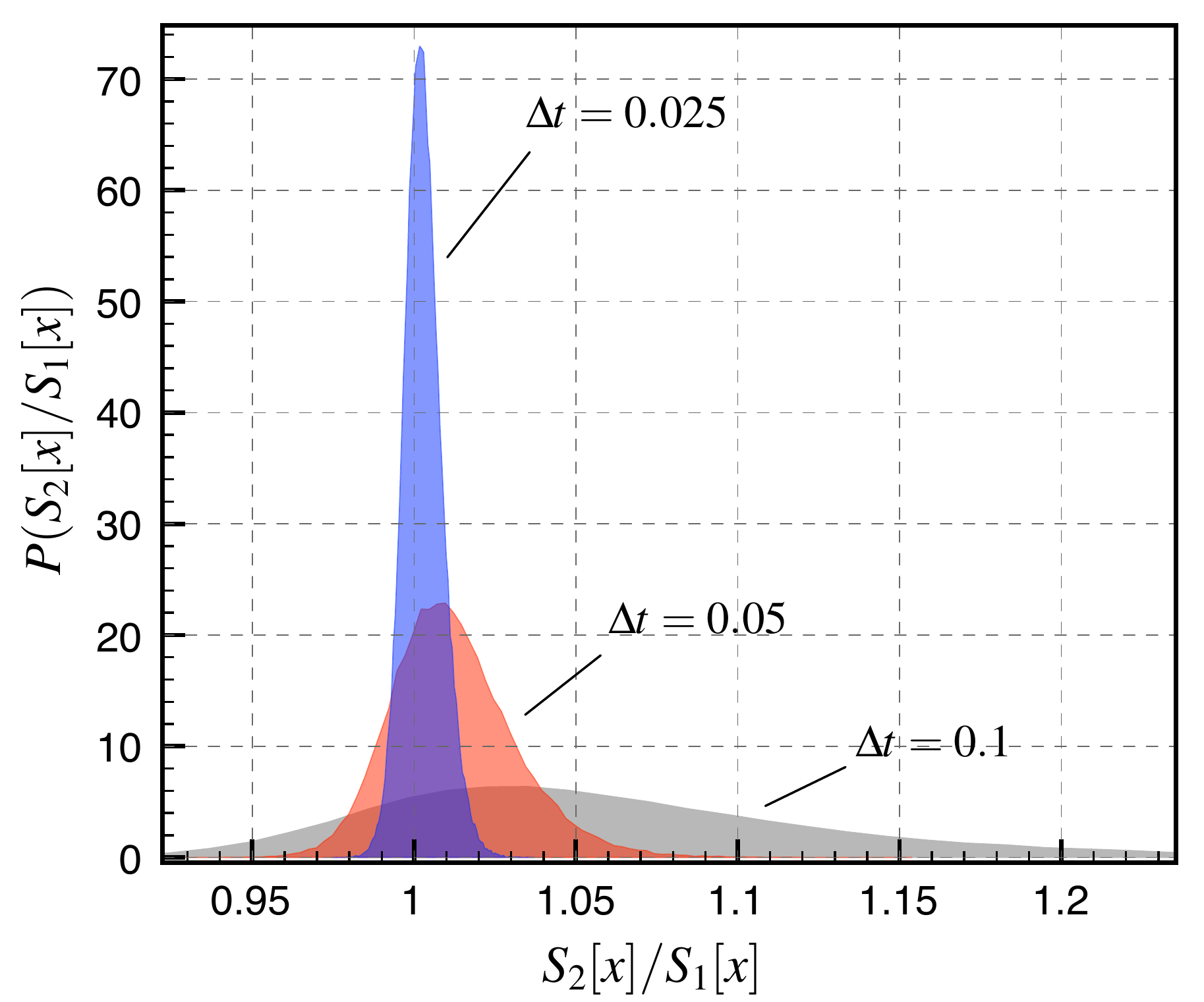} 
\caption{Histograms of the ratio $S_2[x]/S_1[x]$ (unnormalized), illustrating the probabilistic equivalence of the actions in the continuum limit. Each histogram was evaluated over 100,000 Brownian trajectories of duration $t = 10$ generated according to Eq.~(\ref{ito}) for three different $\Delta t$ (indicated in the figure), using the quartic potential $U(x) = x^4/4$ and the parameters $D = 1$, $\zeta = 1$.
}
\label{fig:PS2S1}
\end{center}
\end{figure}

These observations are summarized in Figures~\ref{fig:S1S2traj} and \ref{fig:PS2S1}.

\section{Applications} \label{sec:applications}

The fact that the relative probability of observing a given diffusive path is given by a Boltzmann-like factor (cf. Eq.~(\ref{path-int})) shows that the action and the diffusion constant are for trajectories what the energy and the temperature are for conformations in the canonical ensemble. This analogy has led many authors to translate equilibrium methods (e.g. sampling, reweighting and minimization techniques), to the context of diffusive trajectories.\cite{pratt86,dellago98a,doniach01,jarzynski99-NucPhys,zuckerman99,zuckerman00,xing06,dellago98b,elber00,orland06} In this section, I will argue that the ensemble of trajectories thus generated (or reweighted) is insensitive to the choice between Eq.~(\ref{S1}) or Eq.~(\ref{S2}), while most probable trajectories are more accurately obtained by minimizing $S_2$ instead of $S_1$.

\subsection{Reweighting}

If one is given an ensemble of diffusive trajectories generated under some conditions (e.g. for some potential $U(x)$ and diffusion constant $D$), it is in principle possible to ``reweight'' them so as to obtain a new ensemble corresponding to different conditions (e.g. a different potential $U^*(x)$, with same diffusion constant).\cite{jarzynski99-NucPhys,zuckerman99} In this example, the reweighting factor is given by
\begin{equation} \label{reweight}
  \frac{P^*[\{x_i\}|x_0]}{P[\{x_i\}|x_0]} = e^{-\Delta S/2D},
\end{equation}
where $\Delta S = S^*[x_0 \cdots x_N] - S[x_0 \cdots x_N]$ is the action difference due to the new potential $U^*(x)$, and $D$ is the diffusion constant associated with the given ensemble of trajectories. The notation $P[\{x_i\}|x_0]$ stands for the probability of observing a trajectory $\{x_i\} \equiv x_1 \cdots x_N$ given the initial point $x_0$. The results of the previous section immediately indicate that the choice between $S_1$ or $S_2$ for $S$ is immaterial for this expression: Since the actions are equal (in a probabilistic sense) when evaluated over this ensemble, the reweighting factor is effectively the same whether one uses the former or the latter action. 

\begin{figure}
\begin{center}
\includegraphics[width=240pt]{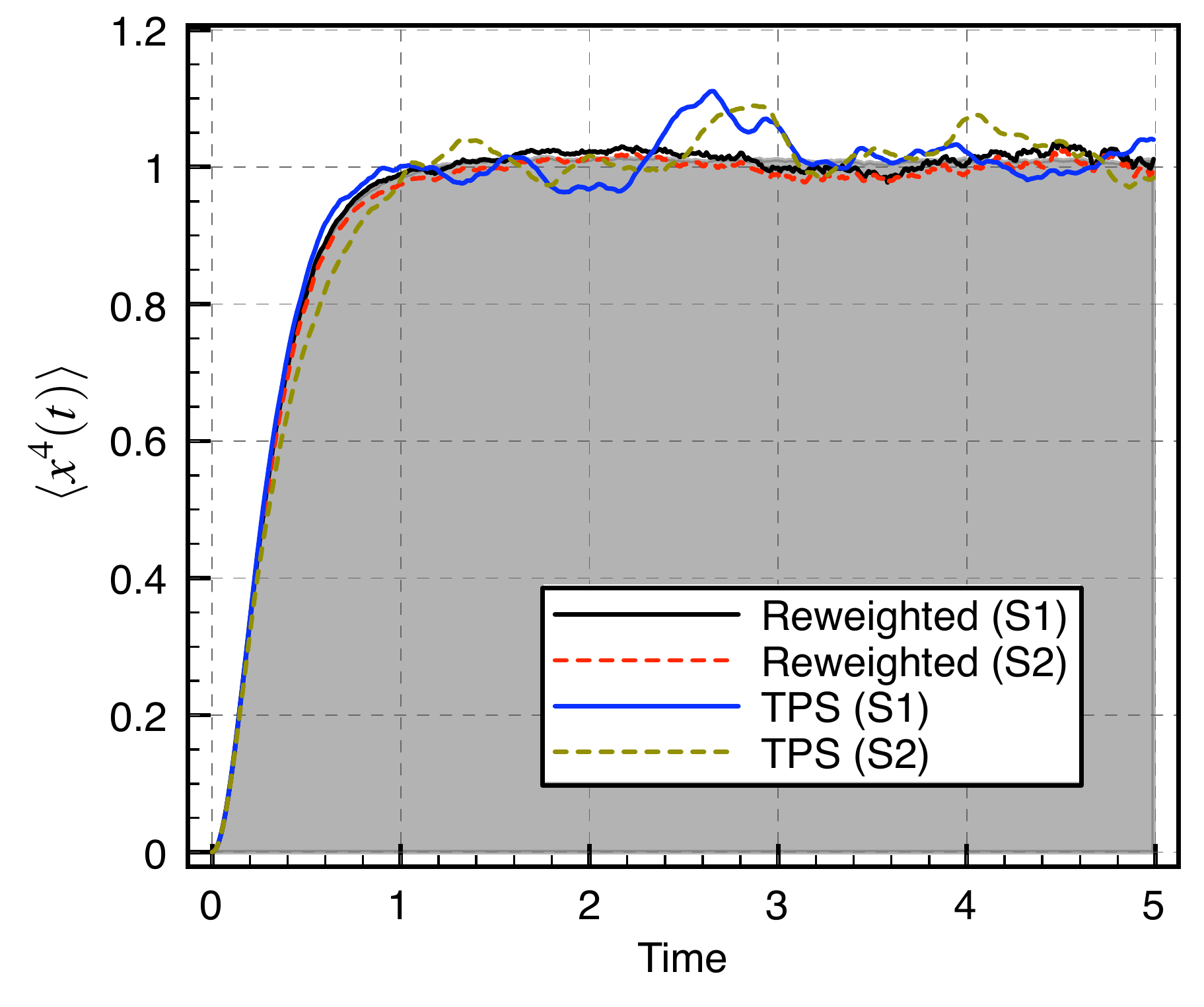} 
\caption{The time-dependent moment $\langle x^4 (t) \rangle$ for a Brownian particle starting at the origin and moving on the quartic potential $U^*(x)=x^4/4$. The shaded gray curve is the result of a direct simulation using Eq.~(\ref{ito}). The curves with smaller fluctuations about $\langle x^4(t) \rangle = 1$ are the results obtained by reweighting $10^5$ free ($U(x)=0$) Brownian trajectories according to Eq.~(\ref{reweight2}), using both $S_1$ (black solid curve) and $S_2$ (red dashed curve). The remaining curves correspond to a one-ended boundary condition version of the ``transition path sampling'' (TPS) Monte Carlo algorithm,\cite{pratt86,dellago98a} with relative probability given by Eq.~(\ref{sampleratio}) and action given by $S_1$ (blue solid curve) and $S_2$ (green dashed curve). The total number of trajectories used for the TPS algorithm was $5\times 10^6$. (The noise level for the TPS results is high even for such a large number of samples because the present Monte Carlo algorithm has not been optimized for efficient trajectory sampling). The parameters adopted throughout are $\Delta t = 0.01$, $\zeta = 1$, and $D=1$. For these values of $D$ and $\zeta$, the exact asymptotic result (e.g. via the virial theorem) is unity.
}
\label{fig:x4}
\end{center}
\end{figure}

Figure~\ref{fig:x4} presents a simple example of this observation, where free trajectories ($U(x)=0$) were reweighted so as to give an ensemble corresponding to a particle subject to a quartic potential $U^*(x) = x^4/4$, using the formula
\begin{equation} \label{reweight2}
  \lang x^4(t) \rang_{U^*} = \lang e^{-\Delta S/2D} x^4(t) \rang_U,
\end{equation}
where the specific functional form of $\Delta S$ depends on the action adopted, and the particle always starts at the origin for both averages. 

\subsection{Sampling} \label{sec:sampling}

The case of sampling is subtler. Since now the actions {\em generate} the ensemble of trajectories itself, one can no longer straightforwardly rely on the above argument, which assumed the ensemble of trajectories was given. Here one is sampling the entire space of trajectories with weight dictated by the action,\cite{pratt86,dellago98a} i.e.
\begin{equation} \label{sampleratio}
  P[\{x_i\}|x_0] \propto e^{-S[x_0 \cdots x_N]/2D},
\end{equation}
where $D$ is the desired diffusion constant. The question remains as to whether the ensemble thus generated is sensitive to the choice between $S_1$ and $S_2$. 

A short (but not particularly illuminating) answer can be given as follows. Since the variable $x$ is Markovian (cf. Eq.~(\ref{langevin})), the statistical properties of its trajectories are fully specified by its propagator, $P(x|x_0;t)$. The problem of equivalence of trajectory ensembles thus reduces to that of equivalence of propagators. When one adopts $S_1$ as the action, one is sampling exactly the same ensemble of trajectories as that generated by Eq.~(\ref{ito}). In the continuum limit, it is known that the propagator corresponding to this Ito-like equation of motion solves the Fokker-Planck equation\cite{gardiner-book}
\begin{equation} \label{FP}
  \frac{\partial P}{\partial t} = -\frac{\partial}{\partial x} \left( \frac{F}{\zeta} P\right) + D \frac{\partial^2 P}{\partial x^2}.
\end{equation}
But this is precisely the same Fokker-Planck equation solved by the path integral in Eqs.~(\ref{path-int})-(\ref{S}),\cite{hunt-ross81} which defines the propagator of the ensemble of trajectories generated by $S_2$ in the continuum limit; i.e. in this limit, the propagators corresponding to the ensembles of trajectories generated by $S_1$ and $S_2$ are identical. Thus, by the Markovian argument above, the two ensembles are statistically equivalent.

A more constructive proof of equivalence can also be given by recalling that, effectively, $S_1$ and $S_2$ only differ when the trajectories over which they are evaluated have a diffusion constant other than that entering the definition of $S_2$. Thus, what one needs to show is that the probability of sampling a path with diffusion constant different from the one specified in Eq.~(\ref{sampleratio}) is vanishingly small, whether one uses $S_1$ or $S_2$; for the remaining trajectories whose sampling probability is in principle non-zero, the actions are effectively equal and hence lead to the same relative weight between the trajectories. 

To proceed, notice that both actions contain a Wiener (free diffusion) term, and thus -- irrespective of the action chosen -- Eq.~(\ref{sampleratio}) can be factored as
\begin{equation}
  P[\{x_i\}|x_0] \propto P_\text{free}^{(D)}[\{x_i\}|x_0] \cdot P_\text{int}[\{x_i\}|x_0],
\end{equation}
where
\begin{equation} \label{Pfree}
  P_{\text{free}}^{(D)}[\{x_i\}|x_0] \propto e^{ - \sum_{n=0}^{N-1} \frac{(x_{n+1}-x_n)^2}{4 D \Delta t} },
\end{equation}
and $P_\text{int}$ is an action-specific distribution accounting for the interaction of the particle with the potential $U(x)$. The proof is finished by showing that the presence of the modulating factor $P_{\text{free}}^{(D)}$ leads to a sharply peaked (i.e. zero variance) distribution of diffusion constants about $D$ (cf. Appendix), which effectively rules out the sampling of trajectories that violate the equivalence between $S_1$ and $S_2$.

The above observations are illustrated in Figure~\ref{fig:x4} again for a particle moving in the quartic potential $U(x)=x^4/4$. The sampling method adopted is a simple extension of the transition path sampling algorithm\cite{pratt86,dellago98a} where only the initial point $x(0)=0$ is fixed. The method uses a straightforward Monte Carlo procedure, where new trajectories are attempted by displacing one of the various $x_i$ by a given amount, and accepting or rejecting the new trajectory according to the standard Metropolis rule\cite{allen-tildesley87} (recall the analogy between action-diffusivity and energy-temperature above).

\subsection{Minimization and most probable trajectories} \label{sec:min}

What has been shown so far is that in trajectory space, the surfaces defined by $S_1$ and $S_2$ are nearly coincident over the overwhelming majority of trajectories representative of the desired ensemble. Nonetheless, appreciable discrepancies between these two surfaces arise when one considers trajectories with ``wrong'' diffusivity; in particular, the actions generally lead to entirely different values when evaluated over differentiable functions (Sec.~\ref{sec:equivalence}). As we shall see below, this bears important consequences to the study of most probable trajectories.

Since the probability of observing a given trajectory decreases exponentially fast with its action (cf. Eq.~(\ref{sampleratio})), the trajectories with greatest statistical weight are those for which the action is minimal. For the moment, this will be our operational definition of ``most probable trajectory.'' Consider therefore the problem of minimizing $S_1$ and $S_2$ in the space of all possible trajectories $x_0 \cdots x_N$. It is easy to see that the minima of both actions correspond to differentiable functions in the continuum limit. Indeed, because of the free diffusion contribution common to both actions, namely
\begin{equation}
  S_\text{free}[x_0 \cdots x_N] = \frac{1}{2} \sum_{n=0}^{N-1} \frac{\Delta x_n^2}{\Delta t},
\end{equation}
where $\Delta x_n = x_{n+1} - x_n$, the actions diverge in the continuum limit when evaluated over diffusive trajectories, for which $\Delta x_n^2 \sim \mathcal{O}(\Delta t)$. Differentiable functions, on the other hand, yield a finite $S_\text{free}$ contribution, as in this case $\Delta x_n^2 = \mathcal{O}(\Delta t^2)$ (i.e. $\dot{x}$ is finite). Thus, we arrive at the following situation of conflict concerning most probable trajectories: Not only are they highly atypical of diffusive processes (insofar as they are differentiable), but also their very identity is sensitive to the choice between $S_1$ and $S_2$ (as these actions are generally different for differentiable paths). The story of minimization is thus qualitatively different from the case of reweighting and sampling.

The answer to the first conflict is relatively easily settled, and as we shall see hints at the solution of the second. The phenomenon associated with it is not unlike that observed in equilibrium sampling studies: Although the most likely conformations of a given system (e.g. atomic clusters, proteins, liquids, etc.) in principle correspond to minima in the energy landscape, in practice thermal fluctuations will keep the system near to, but hardly in the precise location of the minima themselves. Thus, although differentiable paths are virtually never observed in diffusive problems, they can provide the motif upon which fluctuations -- i.e. diffusive trajectories -- emerge. This in turn reveals the importance of the neighborhood of the minima in such analyses, which was neglected in the above discussion of most probable trajectories.

\begin{figure}
\begin{center}
\includegraphics[width=240pt]{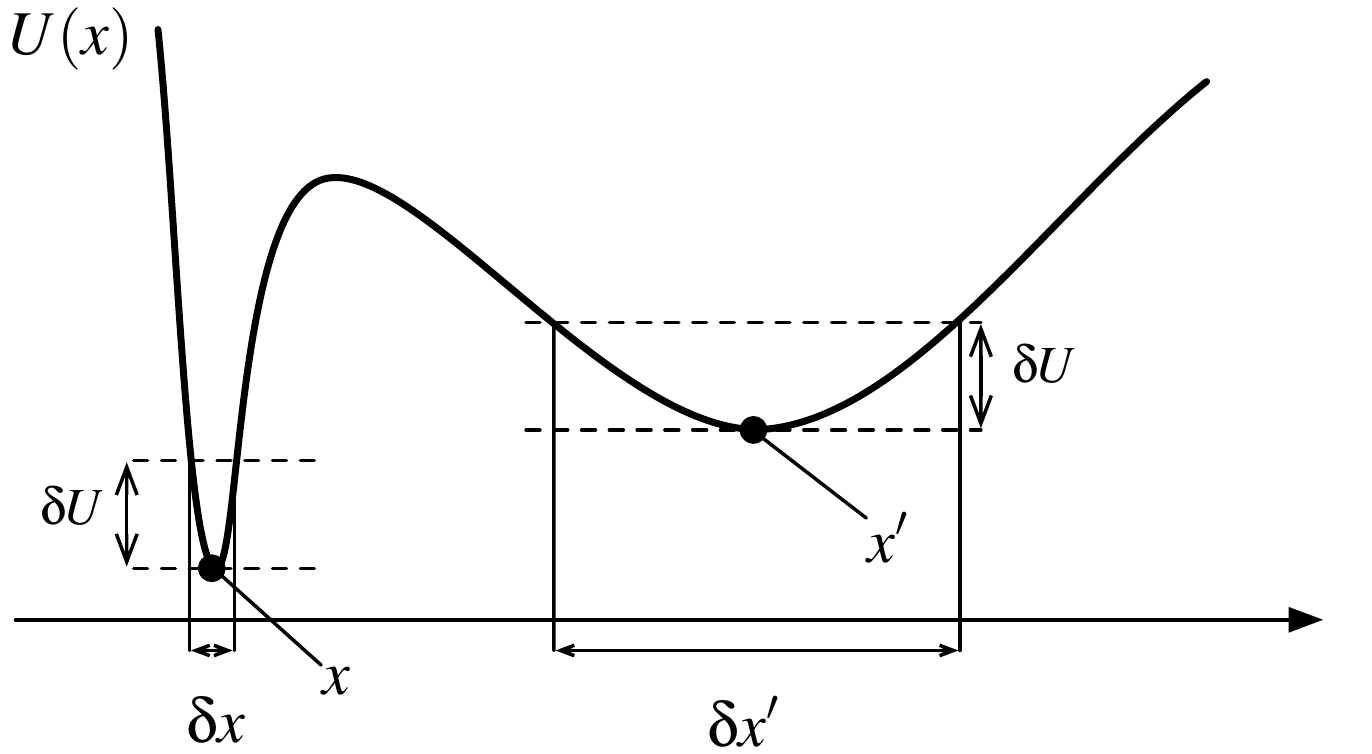} 
\caption{Simple illustration of the entropic role played by the curvature near conformational minima. For the same amount of thermal energy, say $\delta U$, the typical amplitude $\delta x$ of conformational fluctuations near $x$ is much smaller than that near $x'$, $\delta x'$ (use e.g. $\delta U \approx \frac{1}{2} U''(x) \delta x^2$). Thus, although conformations in the neighborhood of $x$ are energetically favored over those near $x'$, the opposite is true when one considers the entropy due to the different curvatures, and the final relative population of $x$- and $x'$-``like'' conformations should embody both contributions, e.g. $P(x') \delta x' / P(x) \delta x = [ P(x') / P(x) ] \times [U''(x')/U''(x)]^{-1/2}$ rather than simply $P(x')/P(x)$.
}
\label{fig:U}
\end{center}
\end{figure}

An illustration of how the nature of such neighborhoods can affect the definition of most probable trajectories is given in Fig.~\ref{fig:U} for a simple one-dimensional potential. By analogy, we are thus led to look at relative probabilities of the type
\begin{equation} \label{relprob}
  \frac{ P(\{ x'_i \}|x_0) \, \delta x'_1 \cdots \delta x'_N}{ P(\{ x_i \} |x_0) \, \delta x_1 \cdots \delta x_N} = e^{-[\Delta S - 2D \Delta \Sigma]/2D},
\end{equation}
where $\Delta S = S[x'_0\cdots x'_N] - S[x_0\cdots x_N]$ and similarly for the ``trajectory entropy'' difference $\Delta \Sigma$, where
\begin{equation} \label{sigma}
  \Sigma[x_1\cdots x_N] \equiv \ln \delta x_1 \cdots \delta x_N.
\end{equation}
Of course, the definition of $\Sigma$ depends on the definition of the measure $\delta x_1 \cdots \delta x_N$, which is not unique; essentially, it depends on how far and in what ways one is willing to accept points away from a given motif trajectory as mere fluctuations about it. Perhaps the simplest definition is obtained by looking at the fluctuations of each individual $x_n$ for a given typical action fluctuation $\delta S$, i.e. by inverting the expansion
\begin{equation}
  \delta S = \frac{\partial S}{\partial x_n} \delta x_n + \frac{1}{2} \frac{\partial^2 S}{\partial x_n^2} \delta x_n^2 + \ldots.
\end{equation}
Assuming one is dealing with local minima or points of very low gradients in comparison to curvatures, a second-order truncation yields the simple measure
\begin{equation} \label{measure}
  \delta x_1 \cdots \delta x_N = \sqrt{ \frac{2 \delta S}{S''(x_1)} } \cdots \sqrt{ \frac{2 \delta S}{S''(x_N)} },
\end{equation}
where $S''(x_n)$ is a shorthand for $\partial^2 S[x_0\cdots x_N]/\partial x_n^2$. I will now proceed to show that with this simple choice of measure in Eq.~(\ref{relprob}), the second conflict above is eliminated.

According to the new relative probability in Eq.~(\ref{relprob}), in order to find most likely trajectories the quantity to be minimized is not the action alone, but rather the ``trajectory free energy''
\begin{equation}
  \mathcal{F}[x_0 \cdots x_N] \equiv S[x_0 \cdots x_N] - 2D \, \Sigma[x_1 \cdots x_N].
\end{equation}
Because the actions diverge for diffusive trajectories, any minimization process will quickly be restricted to what would be differentiable trajectories in the continuum limit. Let us therefore evaluate the action derivatives required by $\Sigma$ (cf. Eqs.~(\ref{sigma}) and (\ref{measure})) over such trajectories. From the expressions in Eqs.~(\ref{S1}) and (\ref{S2}), up to terms explicitly of first order in $\Delta t$ one gets
\begin{align}
  \frac{\Delta t}{2} \frac{\partial S_1}{\partial x_n} & = -(x_{n+1} - 2x_n + x_{n-1}) \\
  & - \frac{\Delta t}{\zeta} \left[ (x_{n+1}-x_n) F'_n - F_n + F_{n-1} \right] \nonumber \\
\frac{\Delta t}{4} \frac{\partial^2 S_1}{\partial x_n^2} & = 1 + \frac{\Delta t}{2 \zeta} \left[ 2 F'_n + (x_{n+1}-x_n) F''_n \right], \label{S1pp}
\end{align}
and
\begin{align}
  \frac{\Delta t}{2} \frac{\partial S_2}{\partial x_n} & = -(x_{n+1} - 2x_n + x_{n-1}) \\
  \frac{\Delta t}{4} \frac{\partial^2 S_2}{\partial x_n^2} & = 1.
\end{align}
A couple of observations are immediately apparent from these expressions. First, the gradients of both $S_1$ and $S_2$ are of order $\mathcal{O}(\Delta t)$ for differentiable trajectories [use $x_{n+1}-2x_n + x_{n-1} = \mathcal{O}(\Delta t^2)$ and $(x_{n+1}-x_n)F'_n - F_n + F_{n-1} = \mathcal{O}(\Delta t^2)$], whereas their curvatures diverge as $\mathcal{O}(1/\Delta t)$. Second, to order $\Delta t$, the relative curvatures of $S_1$ are trajectory-dependent, whereas in the case of $S_2$ such curvatures are constant.

Thanks to the first observation, the definition in Eq.~(\ref{measure}) is valid, and we obtain the trajectory entropy corresponding to $S_1$ by invoking the result in Eq.~(\ref{S1pp})
\begin{align}
  \Sigma_1[x_1\cdots x_N] & = -\frac{1}{2} \ln \prod_{n=1}^N \left[ 1 + \frac{\Delta t}{\zeta} F'_n + \mathcal{O}(\Delta t^2) \right] \nonumber \\
           & = -\frac{\Delta t}{2 \zeta} \sum_{n=1}^N F'_n + \mathcal{O}(N \Delta t^2)
\end{align}
where use was made of $(x_{n+1}-x_n) F''_n = \mathcal{O}(\Delta t)$, and trajectory-independent constants are omitted. An analogous computation of the trajectory-dependent components of $S_2$ reveals that
\begin{equation}
  \Sigma_2[x_1\cdots x_N] = \mathcal{O}(N \Delta t^2),
\end{equation}
which vanishes in the continuum limit; i.e., only $S_1$ leads to a trajectory-dependent entropy. Therefore, when evaluated over differentiable trajectories in the continuum limit, the quantity to be minimized in the case of $S_1$ is the ``free energy''
\begin{equation}
  \mathcal{F}_1[x(s)] = S_1[x(s)] - \frac{D}{\zeta} \int_0^t \! ds \, F',
\end{equation}
whereas in the case of $S_2$ the free energy is simply the action itself, i.e.
\begin{equation}
  \mathcal{F}_2[x(s)] = S_2[x(s)].
\end{equation}
Comparing these results with Eqs.(\ref{S1cont}) and (\ref{S2cont}), we see that when fluctuations are taken into account the quantities to be minimized coincide (i.e. $\mathcal{F}_1 = \mathcal{F}_2$), and consequently the aforementioned ambiguity disappears. The present treatment of most probable trajectories then leads to a functional that agrees with more elaborate mathematical approaches using induced measures in function space.\cite{durr78}

\begin{figure}
\begin{center}
\includegraphics[width=240pt]{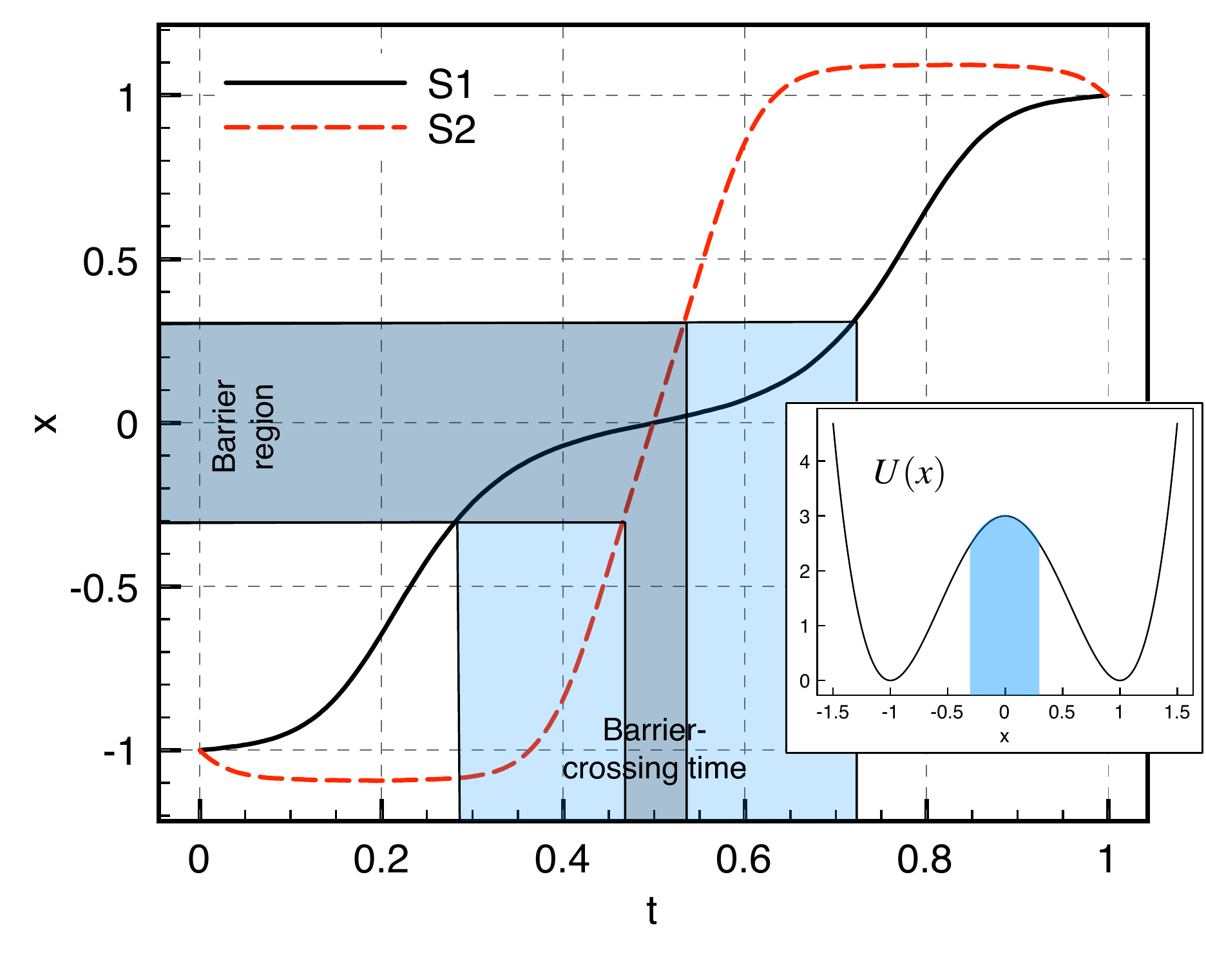}
\caption{The most probable trajectory motifs (``instantons'') connecting the points $x=-1$ and $x=+1$ in one unit of time for the double-well potential Eq.~(\ref{doublewell}) (see inset), according to the minimization of $S_1$ (black solid curve) and $S_2$ (red dashed curve). The parameters are $D=\zeta=1$. Note that without the entropic correction, the actions predict rather different barrier-crossing times; e.g. for the region $-0.3 < x < 0.3$ (shaded area in the $U(x)$ inset) one obtains $\tau_1 \approx 0.43$ and $\tau_2 \approx 0.07$ for $S_1$ and $S_2$, respectively.
}
\label{fig:Instantons}
\end{center}
\end{figure}

\begin{figure}
\begin{center}
\includegraphics[width=240pt]{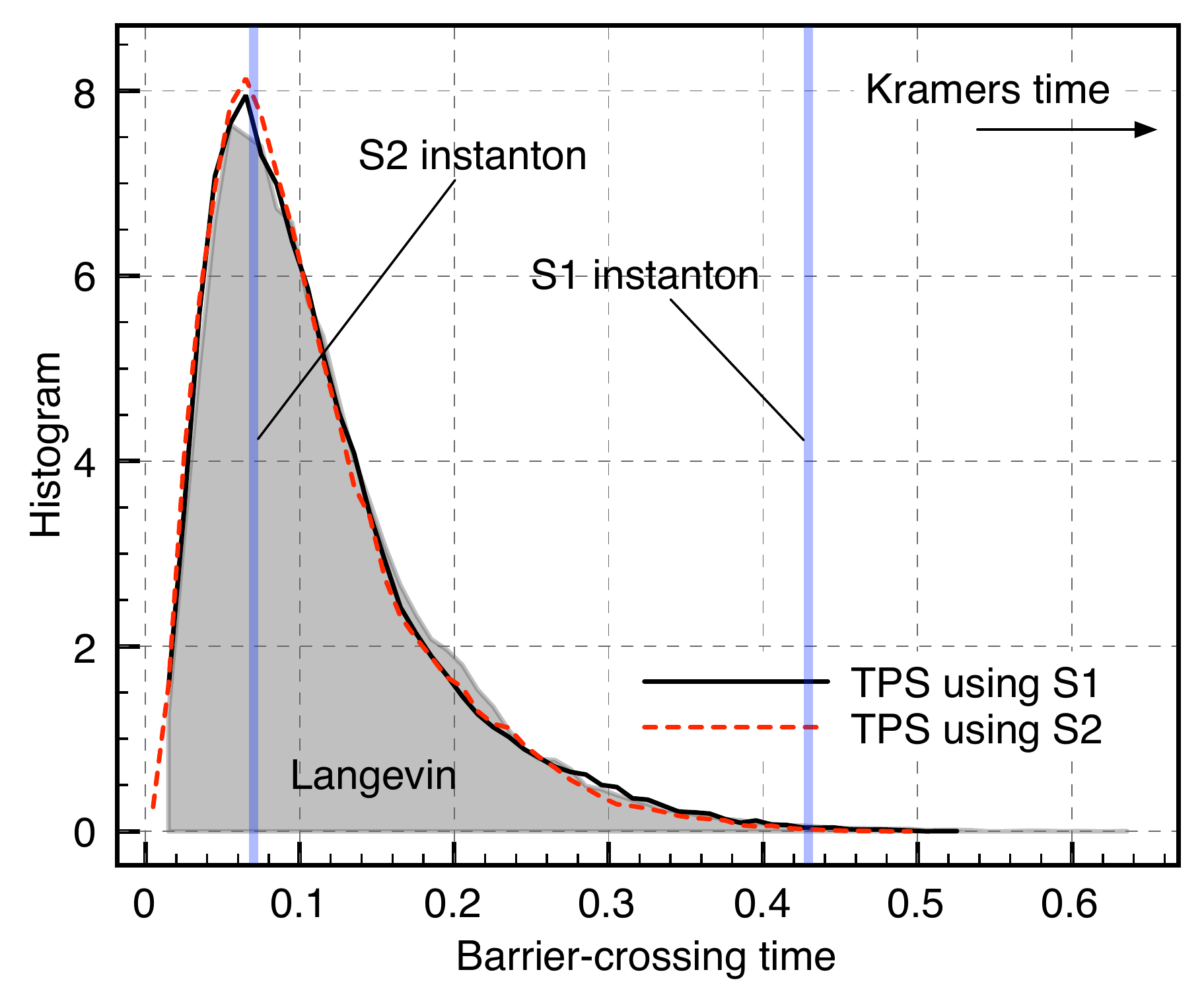}
\caption{Distribution of barrier-crossing times in the barrier region $-0.3 < x < 0.3$ of the double-well potential of Fig.~\ref{fig:Instantons} according to the transition path sampling algorithm (TPS)\cite{pratt86,dellago98a} using both $S_1$ (black solid curve) and $S_2$ (red dashed curve) for a total trajectory length of $t=1$ with $\Delta t = 0.01$ (longer trajectory lengths do not change the shape of the instanton, and hence yield essentially the same barrier-crossing times per instanton). The results from a direct Langevin simulation using Eq.~(\ref{ito}) is also shown (shaded gray curve). In accordance with the discussion of Sec.~\ref{sec:sampling}, the statistics is insensitive to the choice of action. Nonetheless, the predictions based on the minima of the actions alone (vertical blue bars) are quantitatively different (cf. Fig.~\ref{fig:Instantons}). The result based on $S_2$, whose entropic correction is trajectory-independent, is in good agreement with the observed maximum in the distribution. For comparison, the Kramers time $\tau_K \approx 4$ (obtained from the exponential decay of the correlation function $\langle x(0) x(t) \rangle \sim e^{-t/\tau_K}$) was also calculated, and falls well outside the range of the plot.
}
\label{fig:CrossingTimes}
\end{center}
\end{figure}

To illustrate the above observations, consider the one-dimensional problem of a Brownian particle moving according to Eq.~(\ref{langevin}) in the double-well potential
\begin{equation} \label{doublewell}
  U(x) = 3(x^2-1)^2,
\end{equation}
and let us ask what is the most probable trajectory motif that takes the particle from the minimum at $x=-1$ to the minimum at $x=+1$, in one unit of time for $D=\zeta=1$. Figure~\ref{fig:Instantons} shows the predictions obtained by direct minimization of the actions alone. In agreement with our expectation, the actions predict smooth -- i.e. ``differentiable'' -- paths connecting the end-points, with the specific shape depending on whether one uses $S_1$ or $S_2$. In particular, the predicted paths spend rather different times in the barrier region, herein called the ``barrier-crossing time'' (note that this quantity is generally different from the transition, or Kramers' time, which is largely dominated by the residence time in each well). In agreement with our discussion above, however, $S_2$ already embodies the trajectory entropies, and hence its minimum predicts more accurately the overall shape of the most probable trajectories, in particular their barrier-crossing times (cf. Fig.~\ref{fig:CrossingTimes}).

\section{Conclusions}

The present paper was primarily motivated by a conspicuous ambiguity in previous numerical studies of diffusive processes using path integral techniques.\cite{pratt86,dellago98a,doniach01,jarzynski99-NucPhys,zuckerman99,zuckerman00,xing06,dellago98b,elber00,orland06} This ambiguity was introduced in Sec.~\ref{sec:background}, where it was suggested that at least two different actions, namely Eqs.~(\ref{S1}) and (\ref{S2}), are in principle legitimate prescriptions for the discrete representation of the propagator, Eq.~(\ref{path-int}). It was then argued that the actions are indeed equivalent when evaluated over diffusive, non-differentiable trajectories with the same diffusion constant as that appearing in Eq.~(\ref{S2}), while generally leading to different values when evaluated over differentiable paths (Sec.~\ref{sec:equivalence}).

The consequences of these observations to numerical simulations were then explored in Sec.~\ref{sec:applications}, where the equivalence of the actions for the purposes of reweighting and sampling diffusive paths was shown and numerically illustrated, while for problems concerned with the most likely ``motif'' of diffusive trajectories it was argued that $S_2$ already contains the relevant entropic corrections connected with fluctuations about the trajectory motif, and hence should be favored over $S_1$ in direct minimization studies. 

A comforting message in this work is that, contrary to what has been suggested previously,\cite{zuckerman00,orland06} the second-derivative of the potential energy function (cf. Eq.~(\ref{S2})) is not always required for the correct description of diffusive problems (see also Ref.~\onlinecite{predescu07}); it is only if one inquires about most probable trajectories that this numerically inconvenient term plays a role. It is expected that this and the other observations above will aid future authors in making judicious choices for the discrete action in diffusive problems.

\acknowledgments

The author would like to thank Attila Szabo and Gerhard Hummer for numerous discussions, and Gavin Crooks for pertinent suggestions. This research was supported by the Intramural Research Program of the NIH, NIDDK.

\appendix*

\section*{Appendix}

This Appendix contains a derivation that illustrates two results, namely the validity of the Ito formula (Eq.~(\ref{ito2})) for a particular choice of $f(x)$, and the sharply peaked structure of the distribution of diffusivities about $D$ when sampling from $P_\text{free}^{(D)}$.

For the particular choice $f(x) = x^2/2$ in Eq.~(\ref{ito2}), we need to show that, in a mean-square sense,
\begin{equation} \label{ito-app1}
  \sum_{n=0}^{N-1} (x_{n+1} - x_n) x_n - \frac{x_N^2 - x_0^2}{2} + D t = 0
\end{equation}
as $N \to \infty$, $\Delta t \to 0$, with $t = N \Delta t$ constant. The proof offered below follows closely that of Gardiner,\cite{gardiner-book} and proceeds by showing that both the first and the second moments of the left-hand-side of Eq.~(\ref{ito-app1}) are zero for a free Brownian trajectory $\{ x_i \}$ (an extension to more general cases, e.g. trajectories satisfying Eq.~(\ref{ito}), is easily performed by keeping track of the leading order terms in $\Delta t$ only).

The trick is to re-express Eq.~(\ref{ito-app1}) in terms of increments $\Delta x_n \equiv x_{n+1} - x_n$, which for free diffusion are exactly Gaussian-distributed. A simple algebraic manipulation of the sum in Eq.~(\ref{ito-app1}) gives
\begin{align}
  \sum_{n=0}^{N-1} (x_{n+1} - x_n) x_n & = \frac{1}{2} \sum_{n=0}^{N-1} [ x_{n+1}^2 - x_n^2 - \Delta x_n^2 ] \nonumber \\
                                       & = \frac{x_N^2 - x_0^2}{2} - \frac{1}{2} \sum_{n=0}^{N-1} \Delta x_n^2, \label{thesum}
\end{align}
and therefore we need to show that, in the aforementioned limit,
\begin{equation} 
  M_p \equiv \langle ( \ssum_n \Delta x_n^2 - 2Dt )^p \rangle_{x_0} \to 0 \label{app-moments}
\end{equation}
for $p=1,2$, where the average subscript means that the initial point $x_0$ is fixed. The first moment $M_1$ is immediately verified, as $\lang \Delta x_n^2 \rang_{x_0} = 2 D \Delta t$. The second moment gives
\begin{equation}
  M_2 = \lang \ssum_n \Delta x_n^4 \rang_{x_0} + \lang \ssum_{n\neq m} \Delta x_n^2 \Delta x_m^2 \rang_{x_0}
  - (2 D t)^2,
\end{equation}
where the cross-term in the quadratic expansion effectively changed the sign of the last term. Now, for Gaussian variables, $\lang \Delta x_n^4 \rang_{x_0} = 3 \lang \Delta x_n^2 \rang_{x_0}^2$, and since $\Delta x_n$ is uncorrelated with $\Delta x_m$ for $n\neq m$, we get
\begin{align}
  M_2 & = 3N (2 D \Delta t)^2 + N(N-1) (2D \Delta t)^2 - (2Dt)^2 \nonumber \\
      & = 2 (2 D)^2 t \, \Delta t, \nonumber
\end{align}
which vanishes in the continuum limit, thus proving Eq.~(\ref{app-moments}).

As to the second assertion concerning the role of $P_\text{free}^{(D)}$ in fixing the diffusion constant, if one takes as an effective measure of diffusivity of a trajectory the quantity
\begin{equation}
  D_\text{eff}[x_0 \cdots x_N] \equiv \frac{ \sum_{n=0}^{N-1} \Delta x_n^2 } {2t}, 
\end{equation}
one can write Eq.~(\ref{app-moments}) as
\begin{equation} \label{moments-D}
  M_p = \lang ( D_\text{eff}[x_0 \cdots x_N] - D )^p \rang_{x_0} \to 0
\end{equation}
for $p=1,2$, where the continuum limit was defined above. By itself, this result already shows the sharply peaked structure of the distribution of effective diffusivities about $D$. To see where this structure comes from, consider the case of free diffusion. Since in this case the angle-brackets are nothing but averages with respect to the density $P_\text{free}^{(D)}$, and since $P_\text{free}^{(D)}$ depends on the trajectory through $D_\text{eff}$, namely
\begin{equation} \label{PfreeDeff}
  P_\text{free}^{(D)}[\{ x_i\}|x_0] \propto e^{-N D_\text{eff}[x_0\cdots x_N]/2D},
\end{equation}
in analogy with canonical averages one can write
\begin{equation}
  \lang f \rang_{x_0} = \int \! d D' \, \Omega_{x_0}(D') P_\text{free}^{(D)}(D') f(D'),
\end{equation}
where $f$ is an arbitrary function of $D_\text{eff}[x_0 \cdots x_N]$, and 
\begin{equation}
  \Omega_{x_0}(D') = \int \! d\{ x_i \} \, \delta(D' - D_\text{eff}[x_0 \cdots x_N])
\end{equation}
is the ``density of trajectories'' with effective diffusion constant $D'$. With this relation in mind, Eq.~(\ref{moments-D}) shows that the probability density in effective diffusivity space, viz. $\Omega(D') \times P_\text{free}^{(D)}(D')$, is sharply peaked about $D$, as desired. A corollary of this result is that $\Omega$ is a fast-growing function of the effective diffusion constant (indeed, an explicit computation shows that the result is analogous to the microcanonical density of states of an ideal gas, i.e. $\Omega(D) \sim D^{N/2}$ for large $N$).

From a practical standpoint, the above observation is more general than the well known result that the set of differentiable functions ($D_\text{eff}=0$) has measure zero with respect to $P_\text{free}^{(D)}$,\cite{chaichian-book01} as it establishes that the set of {\em all} functions with ``wrong'' diffusion constant (i.e. $D_\text{eff} \neq D$) is also practically precluded from any sampling modulated by $P_\text{free}^{(D)}$.


\begin{thebibliography}{19}
\expandafter\ifx\csname natexlab\endcsname\relax\def\natexlab#1{#1}\fi
\expandafter\ifx\csname bibnamefont\endcsname\relax
  \def\bibnamefont#1{#1}\fi
\expandafter\ifx\csname bibfnamefont\endcsname\relax
  \def\bibfnamefont#1{#1}\fi
\expandafter\ifx\csname citenamefont\endcsname\relax
  \def\citenamefont#1{#1}\fi
\expandafter\ifx\csname url\endcsname\relax
  \def\url#1{\texttt{#1}}\fi
\expandafter\ifx\csname urlprefix\endcsname\relax\def\urlprefix{URL }\fi
\providecommand{\bibinfo}[2]{#2}
\providecommand{\eprint}[2][]{\url{#2}}

\bibitem[{\citenamefont{Zwanzig}(2001)}]{zwanzig01}
\bibinfo{author}{\bibfnamefont{R.}~\bibnamefont{Zwanzig}},
  \emph{\bibinfo{title}{Nonequilibrium Statistical Mechanics}}
  (\bibinfo{publisher}{Oxford Univ. Press}, \bibinfo{address}{New York},
  \bibinfo{year}{2001}).

\bibitem[{\citenamefont{Gardiner}(2004)}]{gardiner-book}
\bibinfo{author}{\bibfnamefont{C.~W.} \bibnamefont{Gardiner}},
  \emph{\bibinfo{title}{Handbook of Stochastic Methods for Physics, Chemistry,
  and the Natural Sciences}} (\bibinfo{publisher}{Springer},
  \bibinfo{address}{Berlin}, \bibinfo{year}{2004}), \bibinfo{edition}{3rd} ed.

\bibitem[{\citenamefont{Hunt and Ross}(1981)}]{hunt-ross81}
\bibinfo{author}{\bibfnamefont{K.~L.~C.} \bibnamefont{Hunt}} \bibnamefont{and}
  \bibinfo{author}{\bibfnamefont{J.}~\bibnamefont{Ross}}, \bibinfo{journal}{J.
  Chem. Phys.} \textbf{\bibinfo{volume}{75}}, \bibinfo{pages}{976}
  (\bibinfo{year}{1981}).

\bibitem[{\citenamefont{Pratt}(1986)}]{pratt86}
\bibinfo{author}{\bibfnamefont{L.~R.} \bibnamefont{Pratt}},
  \bibinfo{journal}{J. Chem. Phys.} \textbf{\bibinfo{volume}{85}},
  \bibinfo{pages}{5045} (\bibinfo{year}{1986}).

\bibitem[{\citenamefont{Dellago
  et~al.}(1998{\natexlab{a}})\citenamefont{Dellago, Bolhuis, Csajka, and
  Chandler}}]{dellago98a}
\bibinfo{author}{\bibfnamefont{C.}~\bibnamefont{Dellago}},
  \bibinfo{author}{\bibfnamefont{P.~G.} \bibnamefont{Bolhuis}},
  \bibinfo{author}{\bibfnamefont{F.~S.} \bibnamefont{Csajka}},
  \bibnamefont{and} \bibinfo{author}{\bibfnamefont{D.}~\bibnamefont{Chandler}},
  \bibinfo{journal}{J. Chem. Phys.} \textbf{\bibinfo{volume}{108}},
  \bibinfo{pages}{1964} (\bibinfo{year}{1998}{\natexlab{a}}).

\bibitem[{\citenamefont{Eastman et~al.}(2001)\citenamefont{Eastman,
  Gronbech-Jensen, and Doniach}}]{doniach01}
\bibinfo{author}{\bibfnamefont{P.}~\bibnamefont{Eastman}},
  \bibinfo{author}{\bibfnamefont{N.}~\bibnamefont{Gronbech-Jensen}},
  \bibnamefont{and} \bibinfo{author}{\bibfnamefont{S.}~\bibnamefont{Doniach}},
  \bibinfo{journal}{J. Chem. Phys.} \textbf{\bibinfo{volume}{113}},
  \bibinfo{pages}{3823} (\bibinfo{year}{2001}).

\bibitem[{\citenamefont{Mazonka et~al.}(1999)\citenamefont{Mazonka, Jarzynski,
  and Blocki}}]{jarzynski99-NucPhys}
\bibinfo{author}{\bibfnamefont{O.}~\bibnamefont{Mazonka}},
  \bibinfo{author}{\bibfnamefont{C.}~\bibnamefont{Jarzynski}},
  \bibnamefont{and} \bibinfo{author}{\bibfnamefont{J.}~\bibnamefont{Blocki}},
  \bibinfo{journal}{Nucl. Phys. A} \textbf{\bibinfo{volume}{641}},
  \bibinfo{pages}{335} (\bibinfo{year}{1999}).

\bibitem[{\citenamefont{Zuckerman and Woolf}(1999)}]{zuckerman99}
\bibinfo{author}{\bibfnamefont{D.~M.} \bibnamefont{Zuckerman}}
  \bibnamefont{and} \bibinfo{author}{\bibfnamefont{T.~B.} \bibnamefont{Woolf}},
  \bibinfo{journal}{J. Chem. Phys.} \textbf{\bibinfo{volume}{111}},
  \bibinfo{pages}{9475} (\bibinfo{year}{1999}).

\bibitem[{\citenamefont{Zuckerman and Woolf}(2000)}]{zuckerman00}
\bibinfo{author}{\bibfnamefont{D.~M.} \bibnamefont{Zuckerman}}
  \bibnamefont{and} \bibinfo{author}{\bibfnamefont{T.~B.} \bibnamefont{Woolf}},
  \bibinfo{journal}{Phys. Rev. E} \textbf{\bibinfo{volume}{63}},
  \bibinfo{pages}{016702} (\bibinfo{year}{2000}).

\bibitem[{\citenamefont{Xing and Andricioaei}(2006)}]{xing06}
\bibinfo{author}{\bibfnamefont{C.}~\bibnamefont{Xing}} \bibnamefont{and}
  \bibinfo{author}{\bibfnamefont{I.}~\bibnamefont{Andricioaei}},
  \bibinfo{journal}{J. Chem. Phys.} \textbf{\bibinfo{volume}{124}},
  \bibinfo{pages}{034110} (\bibinfo{year}{2006}).

\bibitem[{\citenamefont{Dellago
  et~al.}(1998{\natexlab{b}})\citenamefont{Dellago, Bolhuis, and
  Chandler}}]{dellago98b}
\bibinfo{author}{\bibfnamefont{C.}~\bibnamefont{Dellago}},
  \bibinfo{author}{\bibfnamefont{P.~G.} \bibnamefont{Bolhuis}},
  \bibnamefont{and} \bibinfo{author}{\bibfnamefont{D.}~\bibnamefont{Chandler}},
  \bibinfo{journal}{J. Chem. Phys.} \textbf{\bibinfo{volume}{108}},
  \bibinfo{pages}{9236} (\bibinfo{year}{1998}{\natexlab{b}}).

\bibitem[{\citenamefont{Elber and Shalloway}(2000)}]{elber00}
\bibinfo{author}{\bibfnamefont{R.}~\bibnamefont{Elber}} \bibnamefont{and}
  \bibinfo{author}{\bibfnamefont{D.}~\bibnamefont{Shalloway}},
  \bibinfo{journal}{J. Chem. Phys.} \textbf{\bibinfo{volume}{112}},
  \bibinfo{pages}{5539} (\bibinfo{year}{2000}).

\bibitem[{\citenamefont{Faccioli et~al.}(2006)\citenamefont{Faccioli, Sega,
  Pederiva, and Orland}}]{orland06}
\bibinfo{author}{\bibfnamefont{P.}~\bibnamefont{Faccioli}},
  \bibinfo{author}{\bibfnamefont{M.}~\bibnamefont{Sega}},
  \bibinfo{author}{\bibfnamefont{F.}~\bibnamefont{Pederiva}}, \bibnamefont{and}
  \bibinfo{author}{\bibfnamefont{H.}~\bibnamefont{Orland}},
  \bibinfo{journal}{Phys. Rev. Lett.} \textbf{\bibinfo{volume}{97}},
  \bibinfo{pages}{108101} (\bibinfo{year}{2006}).

\bibitem[{\citenamefont{Nitzan}(2006)}]{nitzan-book}
\bibinfo{author}{\bibfnamefont{A.}~\bibnamefont{Nitzan}},
  \emph{\bibinfo{title}{Chemical Dynamics in Condensed Phases: Relaxation,
  Transfer, and Reactions in Condensed Molecular Systems}}
  (\bibinfo{publisher}{Oxford Univ. Press}, \bibinfo{address}{New York},
  \bibinfo{year}{2006}).

\bibitem[{\citenamefont{Chaichian and Demichev}(2001)}]{chaichian-book01}
\bibinfo{author}{\bibfnamefont{M.}~\bibnamefont{Chaichian}} \bibnamefont{and}
  \bibinfo{author}{\bibfnamefont{A.}~\bibnamefont{Demichev}},
  \emph{\bibinfo{title}{Path Integrals in Physics, Volume I: Stochastic Process
  and Quantum Mechanics}} (\bibinfo{publisher}{Institute of Physics},
  \bibinfo{address}{London}, \bibinfo{year}{2001}).

\bibitem[{\citenamefont{Allen and Tildesley}(1987)}]{allen-tildesley87}
\bibinfo{author}{\bibfnamefont{M.~P.} \bibnamefont{Allen}} \bibnamefont{and}
  \bibinfo{author}{\bibfnamefont{D.~J.} \bibnamefont{Tildesley}},
  \emph{\bibinfo{title}{Computer Simulation of Liquids}}
  (\bibinfo{publisher}{Clarendon Press}, \bibinfo{address}{Oxford},
  \bibinfo{year}{1987}).

\bibitem[{\citenamefont{Onsager and Machlup}(1953)}]{onsager53a}
\bibinfo{author}{\bibfnamefont{L.}~\bibnamefont{Onsager}} \bibnamefont{and}
  \bibinfo{author}{\bibfnamefont{S.}~\bibnamefont{Machlup}},
  \bibinfo{journal}{Phys. Rev.} \textbf{\bibinfo{volume}{91}},
  \bibinfo{pages}{1505} (\bibinfo{year}{1953}).

\bibitem[{\citenamefont{{Miller~III} and Predescu}(2007)}]{predescu07}
\bibinfo{author}{\bibfnamefont{T.~F.} \bibnamefont{{Miller~III}}}
  \bibnamefont{and} \bibinfo{author}{\bibfnamefont{C.}~\bibnamefont{Predescu}},
  \bibinfo{journal}{J. Chem. Phys.} \textbf{\bibinfo{volume}{126}},
  \bibinfo{pages}{144102} (\bibinfo{year}{2007}).

\bibitem[{\citenamefont{Durr and Bach}(1978)}]{durr78}
\bibinfo{author}{\bibfnamefont{D.}~\bibnamefont{Durr}} \bibnamefont{and}
  \bibinfo{author}{\bibfnamefont{A.}~\bibnamefont{Bach}},
  \bibinfo{journal}{Comm. Math. Phys.} \textbf{\bibinfo{volume}{60}},
  \bibinfo{pages}{153} (\bibinfo{year}{1978}).

\end{thebibliography}
\end{document}